\documentclass[twocolumn,showpacs,
 aps,prb,superscriptaddress]{revtex4}
\usepackage{graphicx}
\usepackage{epsfig}
\usepackage{wasysym}
\newcommand{\sumnear}{\mathop{\sum}_{\langle i j \rangle}}
\begin{document}
\title{Three-body interactions on a triangular lattice}
\author{Xue-Feng Zhang }
\email{zxf@itp.ac.cn} \affiliation{Institute of Theoretical
Physics, Chinese Academy of Sciences, P.O. Box 2735, Beijing
100190, China} \affiliation{Physics Department and Research Center
OPTIMAS, University of Kaiserslautern, 67663 Kaiserslautern,
Germany}
\author{Yu-Chuan Wen }\affiliation{Center of Theoretical Physics,
Department of Physics, Capital Normal University, Beijing 100048,
China}
\author{Yue Yu }
\affiliation{Institute of Theoretical Physics, Chinese Academy of
Sciences, P.O. Box 2735, Beijing 100190, China}
\date{}

\begin{abstract}
We analyze the hard-core Bose-Hubbard model with both the
three-body and nearest neighbor repulsions on the triangular
lattice. The phase diagram is achieved by means of the
semi-classical approximation and the quantum Monte Carlo
simulation. For a system with only the three-body interactions,
both the supersolid  phase and one third solid disappear while the
two thirds solid stably exists. As the thermal behavior of the
bosons with nearest neighbor repulsion, the solid and the
superfluid undergo the 3-state Potts and the Kosterlitz-Thouless
type phase transitions, respectively. In a system with both the
frustrated nearest neighbor two-body and three-body interactions,
the supersolid and one third solid revive. By tuning the strength
of the three-body interactions, the phase diagram is distorted,
because the one-third solid and the supersolid are suppressed.
\end{abstract}


\maketitle

\section{Introduction}

While it is an adjustable quantum simulator for solving some
difficult quantum problems, such as the high-T$_c$
superconductivity and the fractional quantum Hall effects, the
system of the ultracold molecules trapped in the optical lattice
also provides an ideal toolbox to analyze the general properties
of the quantum many-body systems\cite{review0,review01,review3}.
In the real materials, comparing with the dominant role that the
two-body interactions play, the multi-body interactions are
usually taken as the high-order perturbation. On the other hand,
the man-made Hamiltonian with leading multi-body interactions
exhibits many distinctive phenomenons, such as the non-abelian
topological phases \cite{stringnet} and several novel
phases\cite{ringexchange1,ringexchange2,ringexchange3} originated
from the ring exchange interactions. Recently, because of the
engineering development of the ultracold polar molecules confined
in the optical lattice, the multi-body interactions can be
experimentally realized, especially, the three-body interactions
can be varied in a wide range with the nearest neighbor
interactions from negative to positive \cite{dipole3body2}.

While the dominant three-body interactions result in many exotic
phases\cite{1d3b,squre3b,honey3b,2c3b} on one-dimensional chains
and two-dimensional bipartite lattices, the interplay between the
three-body interactions and the geometry frustration is still
unclear. The frustrated lattices (such as the triangular and
Kagome lattices) manifest themselves by enhancing the quantum
fluctuation and often accompanying with the highly degenerate
ground state. In the magnetic materials with the triangular
structure, the spin liquid which breaks no symmetry has been
observed \cite{tri11}. Meanwhile, the triangular optical lattice,
in which the existence of the supersolid has been numerically
proved\cite{tri1,tri2,tri4,tri6,tri7,tri8,tri9,tri10}, has been
realized by using triple laser
beams\cite{trioptical1,trioptical2}.

\begin{figure}[h]
\includegraphics[width=0.45\textwidth]{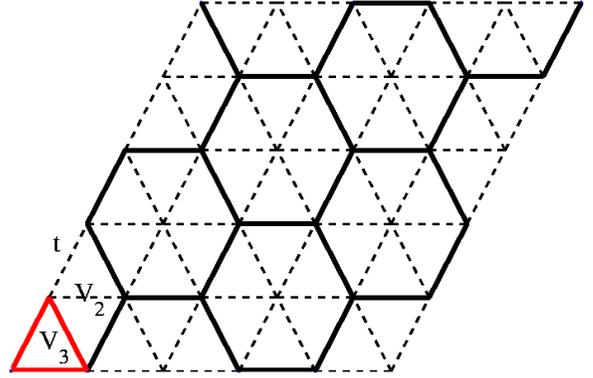}%
\caption{(Color online) The hard-core Bose Hubbard Model on the
triangular lattice. The particles can hop on the bond with
amplitude $t$, and the nearest neighbor repulsion $V_2$ exists
between them. The three-body interactions $V_3$ affect the
particles on the triangle marked by red thick lines. In the
$2/3$($1/3$) solid, the particles(holes) will occupy two
sublattices (black thick line) and form the honeycomb backbone.
\label{lattice}}
\end{figure}

The simplest model to reflect such an interplay is the hard-core
bosons with nearest neighbor two-body and three-body repulsions on
the triangular lattice. The Hamiltonian in the grand canonical
ensemble is shown in Fig.\ref{lattice} and given
\begin{eqnarray}
\nonumber
H&=&-t\sumnear(b_{i}^{\dag}b_{j}+b_{j}^{\dag}b_{i})+V_2\sumnear
n_{i}n_{j} -\mu\mathop{\sum}_in_i\\
&&+V_3\mathop{\sum}_{\langle i,j,k\in\bigtriangleup
\rangle}n_in_jn_k, \label{BH}
\end{eqnarray}
where $b_{i}^{\dag}$ $(b_i)$ is the creation (annihilation)
operator of bosons; $t$ is the hopping parameter; $\mu$ is the
chemical potential; and $V_2$ and $V_3$ are strengthes of the
nearest neighbor two-body and three-body repulsions, respectively,
i.e., $\langle i,j\rangle$ represents $i$ and $j$ are nearest
neighbor sites and $\langle i,j,k\in\bigtriangleup \rangle$ means
$i$, $j$ and $k$ connect with each other two by two and form a
regular triangle. After the Holstein-Primakoff transformation
$b_i^{\dag}\rightarrow S_i^{\dag}$, $b_i\rightarrow S_i$ and
$n_i\rightarrow S_i^Z+1/2$, the bosonic Hamiltonian (\ref{BH}) is
mapped into spin-1/2 XXZ model
\begin{eqnarray}
\nonumber
H&=&-t\sumnear(S_{i}^{\dag}S_{j}+S_{j}^{\dag}S_{i})+(V_2+V_3)\sumnear
S_{i}^Z S_{j}^Z\\
&&-B\mathop{\sum}_iS_i^Z+V_3\mathop{\sum}_{\langle
i,j,k\in\bigtriangleup \rangle}S_i^ZS_j^ZS_k^Z, \label{spinmodel}
\end{eqnarray}
where $B=\mu-3 V_2-3 V_3/2$, and the last term breaks the
particle-hole symmetry.

In this work, we used both the semi-classical approximation
\cite{tri1} and quantum Monte Carlo simulation
\cite{sse1,sse2,clustersse} to study the model described by the
Hamiltonian Eq.~(\ref{spinmodel}). We first assume $V_2=0$ and
merely consider the three-body interactions. We depict the phase
diagram and find the coexistence of both the charge-density wave
order and the bond-ordered wave \cite{1d3b}. Furthermore, at the
finite temperature, the Kosterlitz-Thouless and the 3-state Potts
type phase transitions are found in the superfluid phase and the
solid phase, respectively. After the antiferromagnetic nearest
neighbor interaction is switched on, the phase diagram is derived
in the whole parameter space. For the large three-body
interactions, we observe the separation of the minima of the
superfluid density and the structure factor.

\section{three-body interactions }

In Refs. [\onlinecite{tri1,tri2}], the authors showed that the
frustrated nearest neighbor interaction leads to the charge
density wave (CDW) order. In this section, we show that the
three-body interactions play the same role as the nearest neighbor
interaction does.

\begin{figure}[h]
\includegraphics[width=0.45\textwidth]{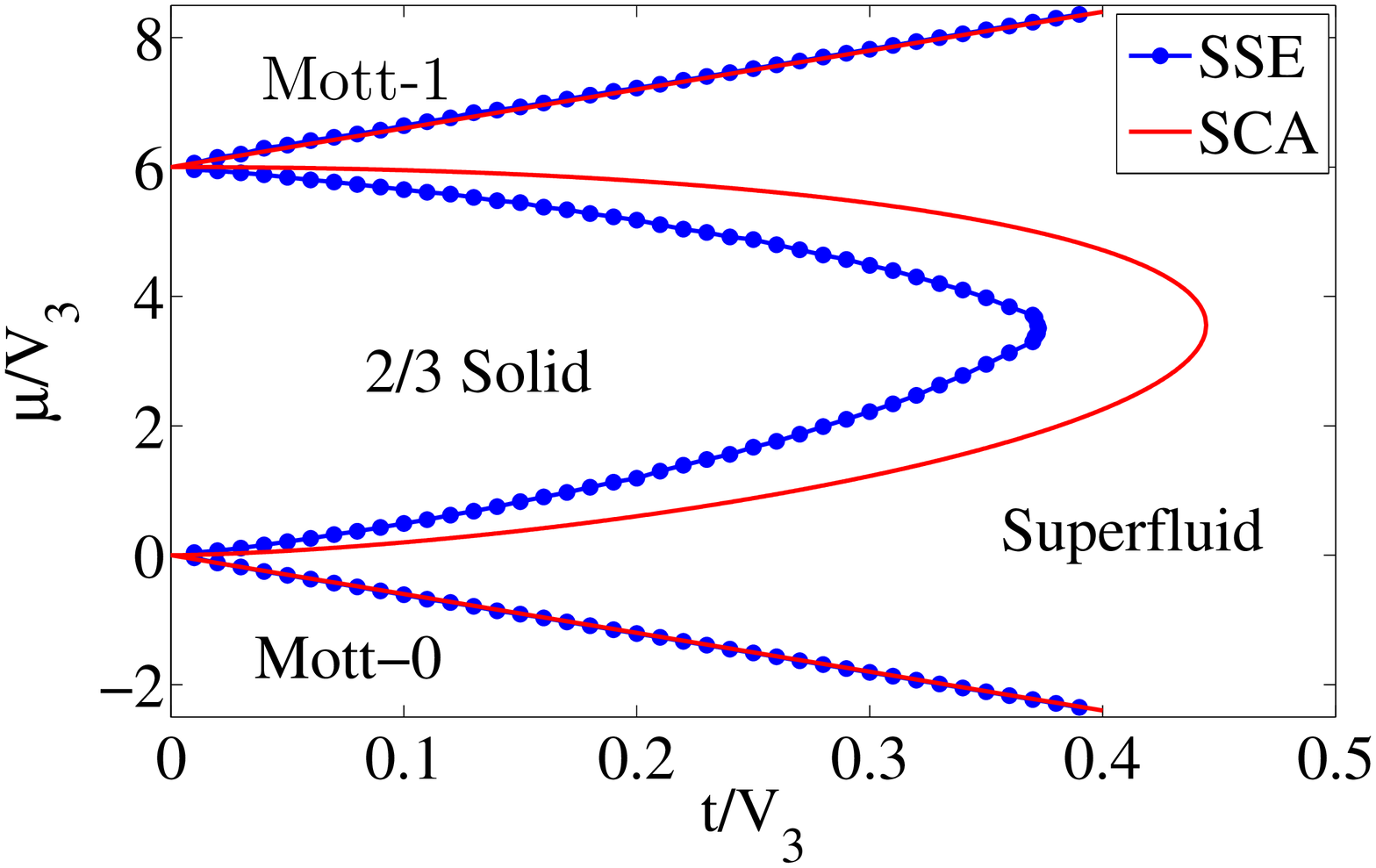}%
\caption{(Color online) The phase diagram with $V_2=0$. The red
line comes from the semi-classical approximation (SCA) and the
dot blue line from quantum Monte Carlo with a stochastic series
expansion (SSE) arithmetic. \label{phasediagram}}
\end{figure}

When the hopping is forbidden, i.e., $t=0$, the ground state at
the absolute zero temperature is exactly known. All the sites are
empty (Mott-$0$ Insulator) when $\mu$ is less then 0. As the
chemical potential increases, a $\sqrt{3}\times\sqrt{3}$ solid
phase appears in order to maximizes the chemical potential part
without costing energy on the three-body repulsions. When
$\mu>6V_3$, the system is fully occupied (Mott-$1$ Insulator),
because the energy paid in the three-body interactions is less
than its gain from the chemical potential. For a finite $t$ there
may be a superfluid phase. Then, the phase diagram is extended to
the finite $t$ and shown in Fig.\ref{phasediagram}.

\subsection{Semi-classical approximation}

By approximating the quantum spin in Eq.~(\ref{spinmodel}) as a
classical unit vector with the magnitude $1/2$, the system is
taken the semi-classical approximation. The ground state is
determined by minimizing the energy per site. Because of the
3-fold rotational and the transitional symmetries, the lattice can
be divided into three sublattices, and the spins are equivalent in
each one.

The superfluid is classically identified by checking whether all
the spins point at the same direction. By using the
semi-classical approximation, we find that the second order
Mott-$0$(Mott-$1$) insulator-superfluid phase transition at
$h=-2\Delta$ ( $h=2 (2+\Delta)$ ), where $\Delta=2t/V_3$ and
$h=2\mu/3V_3$. Meanwhile, the phase transition between solid and
superfluid state is the first order, and the critical lines can
be analytically expressed with
\begin{eqnarray}
h=\frac{16-3\Delta^2}{12}+\frac{c^{1/3}\cos((\theta-2\pi)/3)}6
\label{bsdsf1_v2_eq_0}
\end{eqnarray}
and
\begin{eqnarray}
h=\frac{16-3\Delta^2}{12}+\frac{c^{1/3}\cos(\theta/3)}6,
\label{bsdsf2_v2_eq_0}
\end{eqnarray}
where
\begin{eqnarray*}
a&=&4096-2304\Delta^2-5760\Delta^3-2160\Delta^4\\
&&-216\Delta^5-27\Delta^6 \\
b&=&\sqrt{c^2-a^2}=48\sqrt{6\Delta^3(8-9\Delta)(8+4\Delta+\Delta^2)} \\
c&=&(256-96\Delta^2+48\Delta^3+9\Delta^4)^{3/2} \\
\theta&=&\arccos(a/c).
\end{eqnarray*}
And the summit of the lobe is at $\Delta=8/9$ and $h=64/27$,
which is given by equalizing Eq.~(\ref{bsdsf1_v2_eq_0}) and
Eq.~(\ref{bsdsf2_v2_eq_0}) at $\theta=\pi$. The peak and the
shape of the lobe in the Fig.\ref{phasediagram} also reflect the
breaking of the particle-hole symmetry. Notice that the
semi-classical approximation is exact in the large-S limit and
the result from this approximation is only qualitatively correct.
To depict the precise phase diagram, we use a stochastic series
expansion (SSE) arithmetic in quantum Monte Carlo simulation.

\begin{figure}[h]
\includegraphics[width=0.45\textwidth]{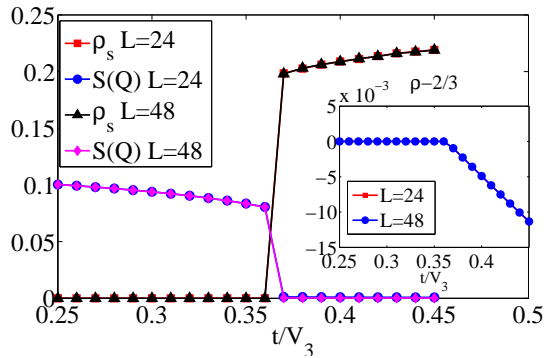}
\caption{(Color online)The structure factor and the superfluid
density (Inset: The density per site) at $\mu/V_3=3.5$ and
$T=0.01V_3$ vs $t/V_3$. $L$ is the linear size of the
lattice.}\label{dtrs_v2_eq_0}
\end{figure}
\subsection{Quantum Monte Carlo}

The cluster SSE \cite{sse1,sse2,clustersse} is taken because of
its high accuracy and efficiency on simulating the system with the
multi-body interactions. The sufficiently low temperature and
large system size ensure the thermodynamic limit to be achieved ,
and the number of the quantum Monte Carlo steps are up to one
million. A solid phase may be described by a charge density wave
(CDW) order parameter, the structure factor $S(\textbf{Q})=\langle
|\mathop{\sum}_{k=1}^N n_k e^{\emph{\textbf{i}} \textbf{Q}
\cdot\textbf{r}_k}|^2\rangle/N^2$ where $N$ is the number of sites
and $\textbf{Q}=(4/3\pi,0)$. Meanwhile, the long range
off-diagonal order is reflected by the finite superfluid density
$\rho_s=\langle W^2/4\beta t\rangle$, where $W$ is the winding
number. In the $2/3$ filling CDW state, the bosons fully occupy
two sublattices, so that the particle and hole on same bond can
partly form a singlet state due to the second order hopping
process. Thus, the bond order wave appears. It is described by the
non-zero bond order structure factor $S_b(\textbf{Q})=\langle
|\mathop{\sum}_{l=1}^{N_b} K_l e^{\emph{\textbf{i}}
\textbf{Q}\cdot \textbf{r}_l}|^2\rangle/N^2$, where $N_b$ is the
number of bonds, $K_l=b^{\dag}_ib_j+b^{\dag}_jb_i$ and $i$, $j$
are two ends of the bond $l$.
\begin{figure}[h]
\includegraphics[width=0.45\textwidth]{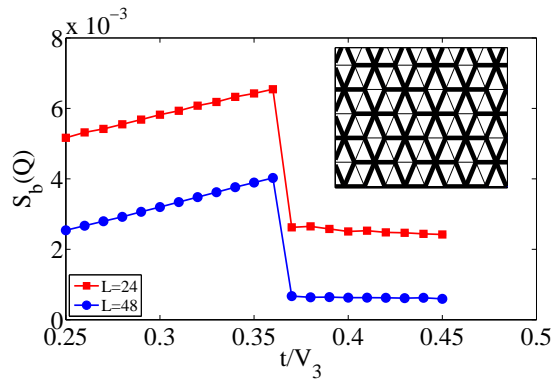}
\caption{(Color online)The $S_b(\textbf{Q})$ at $\mu/V_3=3.5$ and
$T=0.01V_3$ vs $t/V_3$. Inset: the distribution of $K_{i}$ for
$L=12$ in the solid phase. The thick line is $0.24$ and the thin
line is $0.07$.}\label{dtsbq_v2_eq_0}
\end{figure}

\begin{figure}[h]
\includegraphics[width=0.45\textwidth]{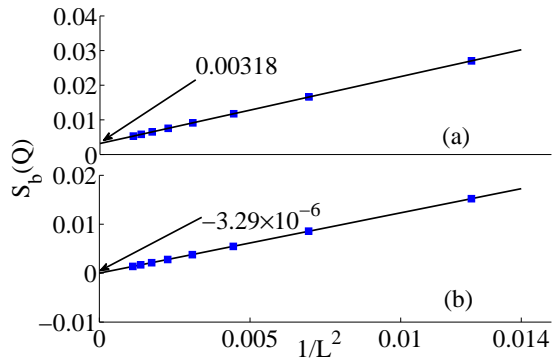}
\caption{(Color online)The finite size scaling of
$S_b(\textbf{Q})$. (a) In the solid phase at $\mu/V_3=3.5$ and
$t/V_3=0.36$. (b) In the superfluid phase at $\mu/V_3=3.5$ and
$t/V_3=1$.}\label{sdfss_v2_eq_0}
\end{figure}

In the inset of the Fig.\ref{dtrs_v2_eq_0}, the density plateau
indicates the incompressible state. And in the
Fig.\ref{dtrs_v2_eq_0}, the finite values of the $S(\textbf{Q})$
and zero superfluid density in the left part (small $t/V_3$)
support the existence of the  solid state ($\rho=2/3$) in this
region. Meanwhile, the finite jump of the superfluid density and
zero $S(\textbf{Q})$ in the right part indicate the first order
superfluid-solid phase transition at the critical point
$t/V=0.36(1)$. The $S_b(\textbf{Q})$ also drops down at the same
critical point, as shown in the Fig.\ref{dtsbq_v2_eq_0}. However,
the dependence on the lattice size requires the finite size
scaling analysis, which is shown in Fig.\ref{sdfss_v2_eq_0}. We
see that the system has a finite bond order wave order in the
solid phase and not in the superfluid. Thus, we confirmed that the
bond order wave and CDW coexist in the solid and do not separate.
However, it is not a novel phase because it can be easily
understood by the local vibrations of the particles or holes. The
inset of the Fig.\ref{dtsbq_v2_eq_0} gives a picture of such an
order, in which the local kinetic energy in unit of $t$ is much
higher on the half filled bonds than the fully filled ones.

The phase diagrams (Fig.\ref{phasediagram}), derived from both the
semi-classical approximation and SSE calculations, matching well,
proves the validity of the semi-classical approximation for such a
model. However, unlike a model with the nearest neighbor repulsive
bosons, neither the $1/3$ filling solid nor the supersolid is
found. The supersolid is in fact the superfluid of the hole
(particle) excitations on the backbone constructed by particles
(holes). Because the bosons are of the hard core, the excited
particle can not hop to the nearest site to have the long range
off-diagonal order. Meanwhile, due to the domain wall
formation\cite{tri2}, the solid order can be destructed by the
infinitesimal density but infinite number of hole excitations, and
the phase transition from the solid to the superfluid is allowed.

\begin{figure}[h]
\includegraphics[width=0.45\textwidth]{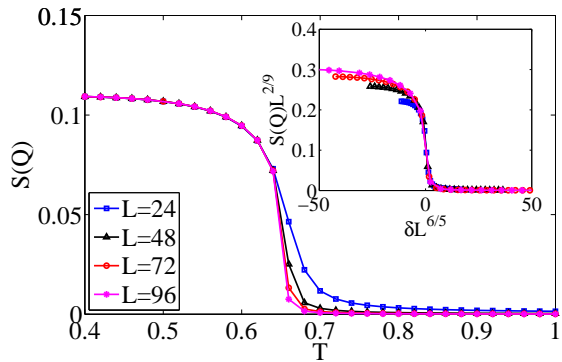}
\caption{(Color online)The structure factor vs temperature for
different sizes at $t/V_3=0.1$ and $\mu/V_3=3$. Inset: the
critical behavior of the 3-states Potts Model universality class
and $\delta=(T-T_c)/t$ with $T_c=0.65$.}\label{ftsolid}
\end{figure}

\subsection{Finite Temperature} In the following, we  study
the finite temperature behaviors in the solid and superfluid
phases. In  Fig.\ref{ftsolid}, we show the thermal melting process
of the CDW order. As the hard-core bosons with only nearest
neighbor interactions on the triangular lattice\cite{tri4}, the
phase transition is expected to be in the universality class of
the 3-state Potts model with the critical exponents $\nu=5/6$ and
$\beta=1/9$. The critical behavior of the structure factor is
$S(\textbf{Q})=f(\delta L^{1/\nu})\times L^{-2\beta},$ where
$\delta=(T-T_c)/t$ and $T_c$ is the fitting critical temperature.
The Fig.\ref{ftsolid} shows the phase transition happens in the
critical point $T_c=0.65$ and also confirms our expectation by the
same function $f$ for different lattice sizes.
\begin{figure}[h]
\includegraphics[width=0.45\textwidth]{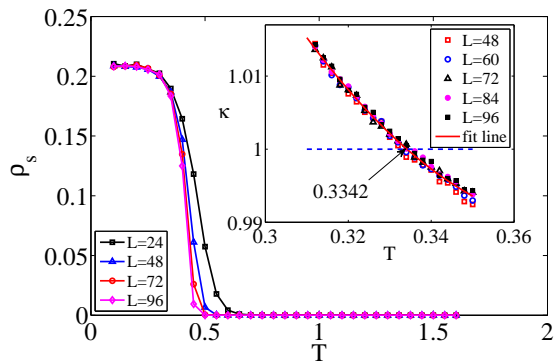}
\caption{(Color online)The superfluid density vs temperature for
different sizes at $t/V_3=0.4$ and $\mu/V_3=3.5$. Inset: The
solution of the Eq.(\ref{kappa}) and the dash line is
$\kappa=1$.}\label{ftsf}
\end{figure}

The superfluid to the normal liquid undergoes the
Kosterlitz-Thouless transition. The critical temperature can be
determined by the renormalization flow and the universal jump of
the superfluid density at $T_c$ (Fig. \ref{ftsf}). The critical
point is given by finding $\kappa=1$, where $\kappa(T)$ is the
integral function
\begin{eqnarray}
  4\ln(L_2/L_1)=\int^{R_1}_{R_2}\frac{dt}{t^2(\ln(t)-\kappa)+t}\label{kappa}
\end{eqnarray}
and $R=3\pi\rho_s/2tT$. We set data of $L=24$ as $R_1$ and the
other sizes as $R_2$, and then plot the $\kappa$ in the inset of
the Fig.\ref{ftsf}. From the Fig.\ref{ftsf}, we observe that the
critical point is around $T_c\approx 0.4$. And the inset shows
the Kosterlitz-Thouless  transition happens at $T_c=0.3342$.
\begin{figure}[h]
\includegraphics[width=0.45\textwidth]{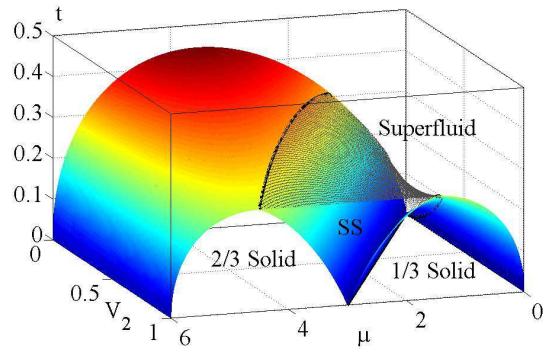}
\caption{(Color online)The phase diagram by SCA with $V_2+V_3=1$.
The solid phases are under the color surface, the supersolid (SS)
exists between the black net and the color surface, and the rest
part is the superfluid. The black lines are the tricritical
lines.}\label{phase2}
\end{figure}

\section{influence on the system with Nearest neighbor interaction}
In the system with only the nearest neighbor interaction, the
$1/3$ filling solid and supersolid phase\cite{tri1,tri2,tri4} can
stably exist. It is intuitively thought that the additional
three-body interactions should have more influences on the $2/3$
CDW order than $1/3$. By using the semi-classical approximation
and setting $V_2+V_3=1$, we plot the phase diagrams
 for different ratios of $V_2$ and $V_3$ in Fig.\ref{phase2}.
 We find that the three-body repulsion strongly enhances the
$2/3$ CDW order, and the $1/3$ CDW order will disappear when
$V_2$ is $0$ which indicates that the $1/3$ CDW order is
associated with the nearest neighbor interaction.
\begin{figure}[h]
\includegraphics[width=0.45\textwidth]{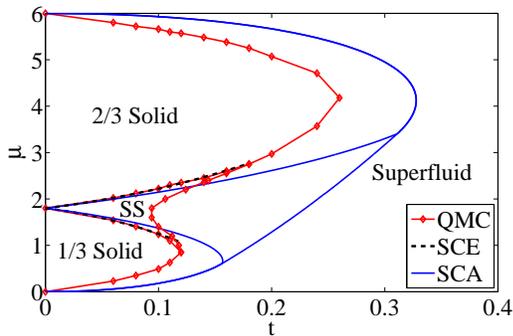}
\caption{(Color online)The phase diagrams at $V_2=0.6$ and
$V_3=0.4$ by using the QMC (red dot), SCA (blue line) and
strong-coupling expansion(SCE) (black dash line).}\label{phase32}
\end{figure}

The strong-coupling expansion  is powerful on solving the
compressible-incompressible second order phase transition.
Therefore, we apply it to the supersolid-solid and Mott-superfluid
phase transitions. The critical line for Mott-1(Mott-0)
insulator-superfluid phase transition is $\mu=6(V_2+V_3)+6t$
($\mu=-6t$), and the $1/3$ and $2/3$ supersolid-solid phase
transition boundaries are given by
\begin{eqnarray}
\nonumber
&&\mu_{1/3}=3V_2-3t-\frac{12t^2}{2V_2+V_3}-\frac{15t^2}{2V_2}-\frac{3t^2}{3V_2+2V_3}\\
\nonumber
&&-\frac{60t^3}{V_2(2V_2+V_3)}-\frac{9t^3}{V_2(3V_2+2V_3)}+\frac{9t^3}{(3V_2+2V_3)^2}\\
&&+\frac{36t^3}{(2V_2+V_3)^2}+\frac{15t^3}{4V_2^2}\label{sd13}
\end{eqnarray}
and
\begin{eqnarray}
\nonumber
&&\mu_{2/3}=3V_2+3t+\frac{12t^2}{2V_2+3V_3}+\frac{15t^2}{2V_2+4V_3}+\frac{3t^2}{3V_2+4V_3}\\
\nonumber
&&+\frac{60t^3}{(V_2+2V_3)(2V_2+3V_3)}+\frac{9t^3}{(V_2+2V_3)(3V_2+4V_3)}\\
&&-\frac{9t^3}{(3V_2+4V_3)^2}-\frac{36t^3}{(2V_2+3V_3)^2}-\frac{15t^3}{4(V_2+2V_3)^2}.\label{sd23}
\end{eqnarray}
When $V_3$ goes to the infinity, the Mott-$1$ insulator will
disappear and the first order phase transition line given by
$\mu_{2/3}$ disappears. In contrary, the $\mu_{1/3}$ is partly
affected in this limit and the Mott-$0$ insulator-superfluid
critical line is not changed. Furthermore, these critical lines
are compared with results from the quantum Monte Carlo and
semi-classical approximation at $V_2=0.6$ and $V_3=0.4$ in
Fig.\ref{phase32}. We see again that the semi-classical
approximation gives a qualitative matched phase diagram to that
comes from the quantum Monte Carlo. In the region that the
strong-coupling expansion is valid, the results fitting with the
data from quantum Monte Carlo are better than those from the
semi-classical approximation.
\begin{figure}[h]
\includegraphics[width=0.45\textwidth]{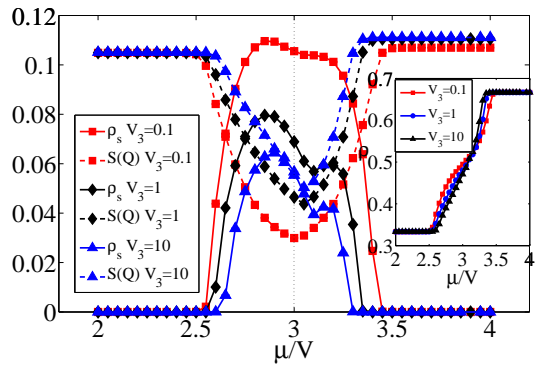}
\caption{(Color online)The superfluid density and the structure
factor at $T=0.02$, $L=24$, $V_2=1$ and $t=0.1$ for different
$V_3$ varying as  $\mu$. Inset: The density per site}\label{nnsr}
\end{figure}

\begin{figure}[h]
\includegraphics[width=0.45\textwidth]{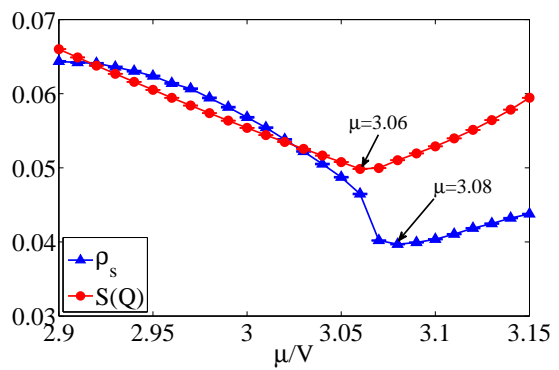}
\caption{(Color online)The superfluid density (blue) and the
structure factor (red) in small region at $T=0.01$, $L=24$,
$V_2=1$ and $t=0.1$ for $V_3=10$ by varying the
$\mu$.}\label{detail}
\end{figure}

In order to detect the three-body interactions' impacts on the
different orders, we use quantum Monte Carlo to simulate several
variables. In terms of  the Fig.\ref{nnsr}, we can find,
comparing to the $1/3$ solid phase which hardly affected by the
three-body repulsion, the
 $2/3$ filling CDW order are strongly enhanced due to decreasing of the local
vibrations. In the supersolid phase, the CDW order is also
enhanced, but the superfluid order becomes weak at the same time.
The influence on the $2/3$ supersolid is stronger than that on the
$1/3$ supersolid, because the repulsive effect is larger in former
case on the superfluid flux. It is also interesting that the
minima of the superfluid density and the structure factor are
separated in the large $V_3$. In the Fig.\ref{detail}, we can
observe it in the small region. Two possible reasons may be used
to interpret it. (i) The minimum of the $S(\textbf{Q})$ determines
the $1/3$-$2/3$ supersolid phase transition, because the
competition between both orders in the critical line may minimize
the $S(\textbf{Q})$. (ii) Because the second order hopping process
in the $2/3$ supersolid are partly prohibited by the three-body
interactions, the holes moving on the honeycomb backbone
constructed by the particles can be approximately treated as, the
quasi-particles forming the superfluid flux on the honeycomb
lattice. For the same reason mentioned in the
Ref.\cite{honeycomb1,honeycomb2}, such dip indicates the geometric
hindrance in the superfluid flow.

\section{Conclusion}
We studied the hard-core Bose-Hubbard Model with nearest neighbor
and three-body repulsions on the triangular lattice by using the
semi-classical approximation and the Quantum Monte Carlo
simulation. In the only three-body repulsions case, we got the
complete phase diagram and find no $1/3$ solid and supersolid
phase exist. By comparing with the CDW and BOW order, we
demonstrate they coexist in the solid phase and the BOW order
trivial results from the local vibrations, so it enlighten us it
needs to be more careful to judge the BOW$+$CDW order. At the
finite temperature, the superfluid phase changes into the normal
liquid phase after a Kosterlitz-Thouless transition. Meanwhile,
the 3-state Potts Model universality class phase transition
emerges when heating up the solid state. And these thermal
properties are same as the solid phase in the system with only
nearest neighbor interaction\cite{tri4}.

After adding the nearest neighbor repulsion, the supersolid and
$1/3$ solid phase revive. The three-body repulsions can enhance
the $2/3$ CDW order and suppress the high order hopping, so they
affect the phase with $2/3$ CDW order harder than $1/3$ CDW order.
However, even up to infinity, it still can not destroy the $2/3$
supersolid, because the excited holes in the honeycomb backbone
constructed by the particles in two sublattices still can form the
superfluid flow without feeling the three-body repulsions. For the
large $V_3$, the minima of the superfluid and CDW order are
separated. The dip of the CDW order may indicate the first order
$1/3$-$2/3$ SS phase transition\cite{tri4}, and the dip of
superfluid density may result from the geometry hindrance.
\section{Acknowledgement}
The authors would like to thank for discussions with S.Eggert and
his suggestions. XFZ thanks the members of SFB/TRR 49. This work
is supported by the National Natural Science Foundation of China,
the national program for basic research of MOST of China, the Key
Lab of Frontiers in Theoretical Physics of CAS, the Natural
Science Foundation of Beijing under Grant No.1092009, the DAAD,
and the DFG via the Research Center Transregio 49. 

\bibliographystyle{apsrev4-1}
\bibliographystyle{apsrev}

\begin{thebibliography}{10}

\bibitem{review0}
Special Issue on Ultracold Polar Molecules: Formation and
Collisions, Eur. Phys. J. D. \textbf{31} (2004).

\bibitem{review01}
A. Micheli, G. K. Brennen and P. Zoller, Nature Physics
\textbf{2}, 341 (2006)

\bibitem{review3}
D. Jaksch and P. Zoller,
\newblock Ann. Phys. \textbf{315}, 52 (2005).

\bibitem{stringnet}
M.A.Levin and X.G.Wen,
\newblock Phy.~Rev.~B {\bf 71}, 045110 (2005)

\bibitem{ringexchange1}
A.W.Sandvik, S. Daul, R. R. P. Singh and D. J. Scalapino
\newblock Phy.~Rev.~Lett. {\bf 89}, 247201 (2002).

\bibitem{ringexchange2}
R.G.Melko, A.W.Sandvik and D.J.Scalapino,
\newblock Phy.~Rev.~B {\bf 69}, 100408(R) (2004).

\bibitem{ringexchange3}
R.G.Melko and A.W.Sandvik,
\newblock Phy.~Rev.~E {\bf 72}, 026702 (2005)

\bibitem{dipole3body2}
H.P.B{\"{u}}chler, A.Micheli and P.Zoller,
\newblock Nature Physics {\bf 3}, 726 (2007)

\bibitem{1d3b}
B.Capogrosso-Sansone S. Wessel, H. P. B{\"{u}}chler, P. Zoller,
and G. Pupillo,
\newblock Phys.~Rev.~B {\bf 79}, 020503(R) (2009)

\bibitem{squre3b}
K. P.Schmidt, J. Dorier and A. M. L{\"{a}}uchli,
\newblock Phys.~Rev.~Lett. {\bf 101}, 150405 (2008)

\bibitem{honey3b}
L. Bonnes, H. P. B\"{u}chler, and S. Wessel,
\newblock New J.~Phys. {\bf 12}, 053027 (2010)

\bibitem{2c3b}
J.K. Pachos and M.B. Plenio,
\newblock Phys.~Rev.~Lett. {\bf 93}, 056402 (2004)

\bibitem{tri11}
Y. Shimizu, K. Miyagawa, K. Kanoda, M. Maesato, and G. Saito
\newblock Phys.~Rev.~Lett. {\bf 91}, 107001 (2003)

\bibitem{tri1}
G.~Murthy, D.~Arovas, and A.~Auerbach,
\newblock Phys.~Rev.~B {\bf 55}, 3104 (1997).

\bibitem{tri2}
S.~Wessel and M.~Troyer,
\newblock Phys.~Rev.~Lett. {\bf 95}, 127205 (2005);
D.~Heidarian and K.~Damle, \newblock Phys.~Rev.~Lett. {\bf 95},
127206 (2005). R.G.~Melko A. Paramekanti, A. A. Burkov, A.
Vishwanath, D. N. Sheng, and L. Balents, \newblock
Phys.~Rev.~Lett. {\bf 95}, 127207 (2005).

\bibitem{tri4}
M.~Boninsegni and N.~Prokof'ev,
\newblock Phys.~Rev.~Lett. {\bf 95}, 237204 (2005).

\bibitem{tri6}
A.~Sen, P.~Dutt, K.~Damle, and R.~Moessner,
\newblock Phys.~Rev.~Lett. {\bf 100}, 147204 (2008).

\bibitem{tri7}
F.~Wang, F.~Pollmann, and A.~Vishwanath,
\newblock Phys.~Rev.~Lett. {\bf 102}, 017203 (2009).


\bibitem{tri8}
H.C.~Jiang M. Q. Weng, Z. Y. Weng, D. N. Sheng, and L. Balents,
\newblock Phys.~Rev.~B {\bf 79}, 020409(R) (2009).


\bibitem{tri9}
D.~Heidarian and A.~Paramekanti,
\newblock Phys.~Rev.~Lett. {\bf 104}, 015301 (2010).

\bibitem{tri10}
L.Pollet, J.D.Picon, H.P.B{\"{u}}chler and M.Troyer,
\newblock Phys.~Rev.~Lett. {\bf 104}, 125302 (2010)

\bibitem{trioptical1}
A.Eckardt P. Hauke, P. Soltan-Panahi, C. Becker, K. Sengstock and
M. Lewenstein,
\newblock EPL {\bf 89}, 10010 (2010)

\bibitem{trioptical2}
C.Becker P. Soltan-Panahi, J. Kronj\"{a}ger, S. D\"{o}rscher, K.
Bongs, K. Sengstock,
\newblock New~J.~Phys. {\bf 12}, 065025 (2010)

\bibitem{sse1}
A.W.~Sandvik,
\newblock Phys.~Rev.~B {\bf 59}, R14157 (1999);

\bibitem{sse2}
O.F.~Sylju{\aa}sen and A.W.~Sandvik,
\newblock Phys.~Rev.~E {\bf 66}, 046701 (2002).


\bibitem{clustersse}
K. Louis and C. Gros,
\newblock Phys.~Rev.~B {\bf 70}, 100410(R) (2004)

\bibitem{honeycomb1}
Stefan Wessel,
\newblock Phys.~Rev.~B {\bf 75}, 174301 (2007)

\bibitem{honeycomb2}
J. Y. Gan, Yu Chuan Wen, Jinwu Ye, Tao Li, Shi-Jie Yang, and Yue
Yu,
\newblock Phys.~Rev.~B {\bf 75}, 214509 (2007)

\end{thebibliography}

\end{document}